\documentclass[a4paper,12pt]{article}
\usepackage{amsmath}
\usepackage{accents}
\usepackage{bm}
\usepackage{dsfont}
\usepackage[english]{babel}
\usepackage{graphicx}
\usepackage[usenames]{color}
\usepackage{colortbl}
\usepackage{mathtools}
\usepackage{commath}
\usepackage[font=footnotesize,figurewithin=none]{caption}
\usepackage[caption=false]{subfig}
\usepackage[toc,page,title]{appendix}
\usepackage{url}

\begin{document}

\centerline {\LARGE{Entanglement of a spin-1/2 Ising-Heisenberg}}
\centerline {\LARGE{diamond spin cluster}}
\centerline {\LARGE{in the thermal bosonic bath}}
\medskip
\centerline {A. R. Kuzmak}
\centerline {\small \it E-Mail: andrijkuzmak@gmail.com}
\medskip
\centerline {\small \it Department for Theoretical Physics, Ivan Franko National University of Lviv,}
\medskip
\centerline {\small \it 12 Drahomanov St., Lviv, UA-79005, Ukraine}

\date{\today}
\begin{abstract}
With the rapid development of quantum information over the last decade, there is a growing need to identify physical systems that can effectively implement quantum computing. One such system is the diamond spin cluster,
which appears in various chemical compounds, including the natural mineral azurite, where copper ions are arranged in this structure. Here, we study the time evolution of a diamond spin cluster with Ising-Heisenberg
interaction under the influence of a thermal bosonic bath, which simulates the environment. Using negativity as a measure, we analyze the entanglement behavior between the central spins of the system.
We demonstrate how the environment influences the presence of entanglement in the system. Specifically, we show that for certain values
of the environment parameters, entanglement increases significantly. Furthermore, we identify the conditions under which entanglement reaches its maximum possible values.
\end{abstract}

\section{Introduction \label{sec1}}

Quantum entanglement is an essential resource for the implementation of quantum information schemes such as quantum teleportation \cite{TELEPORT,Zeilinger1997}, quantum cryptography \cite{Ekert1991},
super-dense coding \cite{Bennett1992}, quantum computing \cite{Nielsen2010,gasparoni2004,Giovannetti20031,Borras2006}, etc. It exists exclusively in quantum systems and arises from the correlations between
their constituent parts \cite{einstein1935,bell1964,aspect982}. To maintain entanglement, it is crucial to protect these systems from environmental disturbances, which can degrade their quality.
At the same time, the physical systems must be easily controllable and measurable. To achieve this, various quantum systems are employed, including nuclear and electronic spins of atoms \cite{phosphorus3,phosphorus1},
superconducting qubits \cite{supcond3}, polarized photons \cite{Zeilinger1997}, trapped ions \cite{SchrodCat1,QSDEGHTI}, ultracold atoms \cite{opticallattice5}, and others.

In the last two decades, bipartite quantum entanglement in diamond spin clusters and chains, both in thermodynamic equilibrium
\cite{bose2005,tribedi2006,ananikian2006,chakhmakhchyan2012,rojas2012,ananikian2012,rojas2014,torrico2016,rojas2017,Zheng2018,Cavalho2019,Ghannadan2022} and in diamond spin clusters evolving under
a magnetic field \cite{kuzmak20231,kuzmak20232}, has been actively studied. In compounds such as ${\rm Ca_3Cu_3(PO_4)_4}$ and ${\rm Sr_3Cu_3(PO_4)_4}$ \cite{drillon1988,drillon1993},
as well as in ${\rm Bi_4Cu_3V_2O_{14}}$ \cite{sakurai2002} and the natural mineral azurite (${\rm Cu_3(CO_3)_2(OH)_2}$) \cite{kikuchi2005}, the ions are arranged in diamond chains.
The electrons between these ions undergo direct exchange interactions concerning both spatial and spin coordinates, forming spin-spin interactions (see, for example, \cite{Eissler2005}).
For instance, in the natural mineral azurite, ${\rm Cu^{2+}}$ ions form a spin-$1/2$ diamond chain, where interactions between spins are governed by the Heisenberg Hamiltonian.
Due to this interaction, the spin states can become entangled, making this system a potential candidate for quantum information applications. It is also worth noting that the thermal bipartite entanglement
of a diamond spin-1 cluster was recently studied in \cite{Cavalho2019}. The authors applied their calculations to a diamond spin-1 cluster formed by ${\rm Ni}^{2+}$ ions in the compound ${\rm [Ni_4(\mu-CO_3)_2(aetpy)_8](ClO_4)}_4$,
where aetpy = 2-aminoethyl-pyridine \cite{escuer1998,hagiwara2006}.

Previous research has focused on studying the effects of temperature and the strength of the magnetic field on entanglement between spins in diamond spin clusters and chains, where the spins remain
in thermodynamic equilibrium. Another study has examined entanglement during the dynamics of a spin diamond cluster in a pure state. In this work, we investigate the evolution of a spin cluster in a bosonic bath,
which induces decoherence. These studies are important because the bath models the photonic or phononic environment, which influences quantum states and their entanglement in real physical systems. Understanding
these effects allows us to identify environmental factors that negatively impact entanglement, enabling their consideration when designing experiments. Thus, we analyze the influence of different bosonic environments
on the behavior of entanglement between central spins in a diamond spin-$1/2$ cluster during its evolution.

The behavior of decoherence in various open quantum systems subjected to different types of noise has been extensively studied. Seminal papers exploring the impact of the environment on both the entanglement and coherence
of composite quantum systems appeared at the beginning of this century \cite{Yu2003,Dodd2004,Yu2004}. These studies demonstrated that, in simple bipartite systems, the effect of weak noise on entanglement is non-additive,
whereas its influence on coherence is additive. This line of research was further developed in the context of a qubit-qutrit system \cite{Ann2008}, where, in addition to confirming earlier results,
the phenomenon of entanglement sudden death was shown to occur for certain quantum states. In recent years, increasing attention has been given to the effects of more general environmental models, particularly
those characterized by Ohmic spectral density \cite{Morozov2012,Chaudhry2013}, on the entanglement dynamics of various spin systems \cite{Ignatyuk2022,Barr2024,Vega2017,Tan2015,Tan2022,Javed2024}.
In our paper, we also use the Ohmic spectral density to describe the environment.

The paper is structured as follows. Section~\ref{modelsec} describes the model of a diamond spin cluster consisting of two central Heisenberg spins and two side Ising spins in a bosonic bath. In Section~\ref{evolsec},
the evolution of the spins is calculated. The expression for negativity, used as a measure of entanglement between the central spins, is derived in Section~\ref{entab}. Results on entanglement behavior under the influence
of different types of bosonic baths are presented in Section~\ref{resultssec}. Finally, conclusions are provided in Section~\ref{conclus}.

\section{Model \label{modelsec}}

\begin{figure}[!!h]
\centerline{\includegraphics[scale=0.60, angle=0.0, clip]{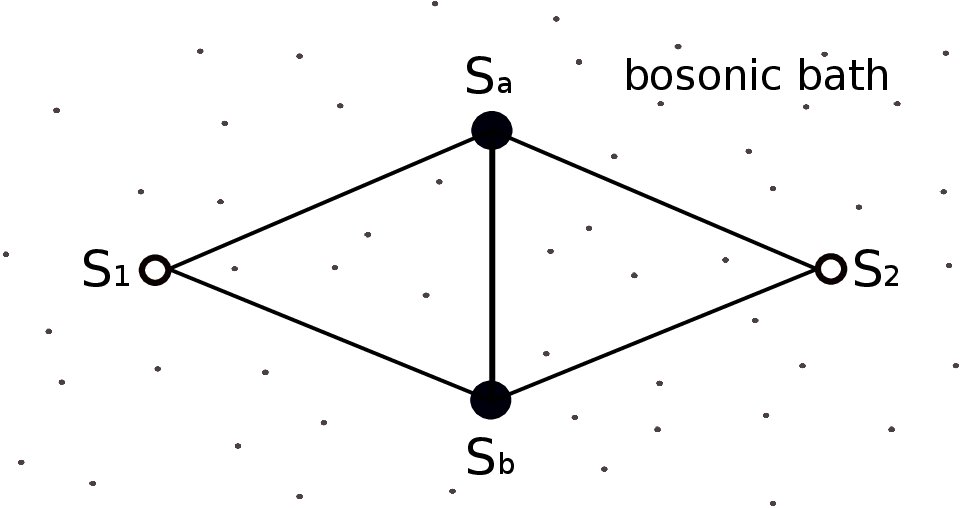}}
\caption{The structure of a diamond spin cluster consisting of two central spins $S_a$, $S_b$ and two side spins $S_1$, $S_2$ in a bosonic environment. Central spins $S_a$ and $S_b$ are described by an anisotropic Heisenberg
Hamiltonian, while their interaction with the side spins $S_1$ and $S_2$ is defined by the Ising model (\ref{spinsystem}). The interaction of spins with bosonic bath is defined by the dephasing model (\ref{spinbath}).}
\label{model}
\end{figure}

The interaction of a diamond spin cluster, consisting of two central spins, $S_a$ and $S_b$, and two side spins, $S_1$, $S_2$, with a bosoonic bath (Fig.~\ref{model}) can be described by the Hamiltonian
\begin{eqnarray}
H=H_{s}+H_{b}+H_{sb},
\label{hamiltonian}
\end{eqnarray}
where the central spins $S_a$ and $S_b$ are governed by an anisotropic Heisenberg Hamiltonian, while their interaction with the side spins $S_1$ and $S_2$ is defined by the Ising model. The Hamiltonian
of the spin subsystem has the form
\begin{eqnarray}
&&H_{s}=H_{ab}+H_{12}+H_{int},\nonumber\\
&&H_{ab}=J\left(S_{a}^xS_{b}^x+S_{a}^yS_{b}^y\right)+J_zS_{a}^zS_{b}^z+h'\left(S_{a}^z+S_{b}^z\right),\nonumber\\
&&H_{12}=h\left(S_{1}^z+S_{2}^z\right),\quad H_{int}=J_0\left(S_{a}^z+S_{b}^z\right)\left(S_{1}^z+S_{2}^z\right),
\label{spinsystem}
\end{eqnarray}
where ${\bf S}_{i}=1/2\left(\sigma_{i}^x{\bf \hat{i}}+\sigma_{i}^y{\bf \hat{j}}+\sigma_{i}^z{\bf \hat{k}}\right)$ represents the spin operator of the $i$-th spin ($i=a,b,1,2$),
$J$, $J_z$ and $J_0$ denote the coupling constants between spins, $h'$, $h$ represent the values of the interaction between spins and an external magnetic field.
The bosonic bath is described by the Hamiltonian consisting of a set of harmonic oscillators with frequencies $\omega_k$ and wave vectors ${\bf k}$
\begin{eqnarray}
H_{b}=\sum_k\omega_k b_k^+b_k,\label{bath}
\end{eqnarray}
where $b_k^+$ and $b_k$ are the creation and annihilation operators of the environment quanta with wave vector ${\bf k}$.
The interaction of spins with a bosonic bath is described by the Hamiltonian, commonly referred to as the dephasing model \cite{Luczka1990},
\begin{eqnarray}
H_{sb}=\left(S_a^z+ S_b^z+ S_1^z+ S_2^z\right)\frac{1}{\sqrt{V}}\sum_k\left(g_k b_k^++g_k^*b_k\right).\label{spinbath}
\end{eqnarray}
Here $V$ is the volume corresponds to the region where the spin-boson subsystem is located, and $g_k$ characterizes the interaction of spins with bosons.
We use the system of units, where the Planck and Boltzmann constants are set to $\hbar =1$, $k_B=1$. It is important to emphasize that spin Hamiltonian $H_s$ commutes with both $H_b$ and $H_{sb}$
\begin{eqnarray}
\left[H_{s},H_{b}\right]=\left[H_{s},H_{sb}\right]=0.\nonumber
\end{eqnarray}
Indeed, all terms in the Hamiltonian $H_s$ (\ref{spinsystem}) contain only spin operators, while the Hamiltonian $H_b$ (\ref{bath}) contains only bosonic operators, which implies that these Hamiltonians commute.
The interaction Hamiltonian $H_{sb}$ (\ref{spinbath}) contains $S_i^z$ operators, which commute with all terms involving the same operators in $H_s$. Additionally operators acting on different spins commute with each other
(for instance, $\left[S_a^xS_b^x,S_1^z+S_2^z\right]=0$). It is also straightforward to show that $\left[S_a^xS_b^x+S_a^yS_b^y,S_a^z+S_b^z\right]=0$.

In the previous papers, we investigated the time-dependent behavior of entanglement \cite{kuzmak20231} and the preparation of entangled states \cite{kuzmak20232} in a spin cluster described by Hamiltonian (\ref{spinsystem}).
Since the spin Hamiltonian $H_s$ mutually commute with $H_b$ and $H_{sb}$ Hamiltonians, we can easily calculate its eigenstates and eigenvalues (Appendix \ref{appegen}).
Let us investigate how the bosonic environment influences the evolution of the diamond spin cluster.

\section{Evolution of the diamond spin cluster \label{evolsec}}

The quantum evolution of the entire system described by the Hamiltonian (\ref{hamiltonian}) can be expressed as follows
\begin{eqnarray}
\rho(t)=e^{-iHt}\rho(0)e^{iHt},
\label{evolution}
\end{eqnarray}
where we assume that the spin and bosonic subsystems in the initial state are separated. The initial state of the spin subsystem can be expressed as $\rho_s(0)=\vert\psi_s(0)\rangle\langle\psi_s(0)\vert$.
The initial state of the bosonic subsystem is in thermodynamic equilibrium, given by $\rho_b(0)=\exp{(-\beta H_b)}/Z_b$, where $Z_b={\rm Tr} \exp{\left(-\beta H_b\right)}$ is the partition function of the bosonic subsystem
and $\beta=1/T$ is the inverse temperature. Since $[H_s,H_b+H_{sb}]=0$, the time-dependent density matrix takes the form
\begin{eqnarray}
\rho(t)=e^{-iH_st}e^{-i(H_b+H_{sb})t}\rho_s(0)e^{-\beta H_b}e^{i(H_b+H_{sb})t}e^{iH_st}/Z_b.
\label{evolution1}
\end{eqnarray}
An arbitrary pure state of four spins can be expressed as follows
\begin{eqnarray}
\vert\psi_s\rangle=\sum_{m_a,m_b,m_1,m_2=\pm 1}c_{m_a,m_b,m_1,m_2}\vert m_a\ m_b\ m_1\ m_2\rangle,
\label{initialstate}
\end{eqnarray}
where $\vert m_i\rangle$ is eigenstate of the $z$-component of the Pauli operator corresponding to the eigenvalue $m_i	=\pm 1$, $c_{m_a,m_b,m_1,m_2}$ are the complex parameters defining the state.
We are interested in the state of the spin subsystem, specifically how the bosonic subsystem affects the spin subsystem.
The density matrix of the spin subsystem is obtained in Appendix~\ref{densitymatrix} by tracing out the bosonic subsystem and has the form
\begin{align}
&\rho_s(t)={\rm Tr}_b \rho(t)=e^{-iH_st}\nonumber\\
&\times\sum_{m_a,m_b,m_1,m_2=\pm 1}\sum_{n_a,n_b,n_1,n_2=\pm 1} c_{m_a,m_b,m_1,m_2}c^*_{n_a,n_b,n_1,n_2}\vert m_a\ m_b\ m_1\ m_2\rangle \langle n_a\ n_b\ n_1\ n_2\vert\nonumber\\
&\times\exp{\left[-\left(\sum_i m_i-\sum_i n_i\right)^2\gamma(t)\right]}\times\exp{\left[-i\left(\left(\sum_i m_i\right)^2-\left(\sum_in_i\right)^2\right)\Delta(t)\right]}\nonumber\\
&\times e^{iH_st},
\label{spinddensitymatrixfinal2}
\end{align}
where we denote the decoherence factors as follows
\begin{eqnarray}
&&\gamma(t)=\sum_k\frac{\vert g_k\vert^2}{4V\omega_k^2}(1-\cos(\omega_k t))\coth(\beta\omega_k/2),\nonumber\\
&&\Delta(t)=\sum_k\frac{\vert g_k\vert^2}{4V\omega_k^2}(\sin(\omega_k t)-\omega_k t).
\label{decfactor}
\end{eqnarray}
Note that the exponential function with $\Delta(t)$ describes the effective indirect interaction between spins mediated by the bosonic bath. While the environment typically induces decoherence in a quantum system,
the spin-boson coupling also gives rise to an additional interaction between the spins. Through the exchange via the bosonic modes, an effective interaction emerges that can generate entanglement
between the spins. In this sense, the bosons act as a medium that couples the spins: one can intuitively imagine a boson becoming entangled with both spins, thereby establishing entanglement between them.
At the same time, this interaction contributes to the growth of entropy within the spin subsystem, leading to decoherence. Since the spin-boson interaction has a pure dephasing character (\ref{spinbath}),
the resulting effective interaction between the spins takes the form of an Ising-type coupling.

Now, using the fact that Hamiltonians $H_{ab}$, $H_{12}$ and $H_{int}$ (\ref{spinsystem}) mutually commute between themselves, we can rewrite density matrix (\ref{spinddensitymatrixfinal2}) as follows
\begin{align}
&\rho_s(t)=e^{-iH_{ab}t}\nonumber\\
&\times\sum_{m_a,m_b,m_1,m_2=\pm 1}\sum_{n_a,n_b,n_1,n_2=\pm 1} c_{m_a,m_b,m_1,m_2}c^*_{n_a,n_b,n_1,n_2}\vert m_a\ m_b\ m_1\ m_2\rangle \langle n_a\ n_b\ n_1\ n_2\vert\nonumber\\
&\times\exp{\left[-i\frac{ht}{2}(m_1+m_2-n_1-n_2)\right]}\nonumber\\
&\times\exp{\left[-i\frac{J_0t}{4}\left((m_a+m_b)(m_1+m_2)-(n_a+n_b)(n_1+n_2)\right)\right]}\nonumber\\
&\times\exp{\left[-\left(\sum_i m_i-\sum_i n_i\right)^2\gamma(t)\right]}\times\exp{\left[-i\left(\left(\sum_i m_i\right)^2-\left(\sum_in_i\right)^2\right)\Delta(t)\right]}\nonumber\\
&\times e^{iH_{ab}t}.
\label{spinddensitymatrixfinal3}
\end{align}
Here we use the eigenequation $\exp{\left(\alpha S_i^z\right)}\vert m_i\rangle=\exp{\left(\alpha m_i/2\right)}\vert m_i\rangle$. To obtain the density matrix of the $S_a$ and $S_b$ spins,
we trace out $\rho_s(t)$ over the $S_1$, $S_2$ spins. As a results, we obtain
\begin{align}
&\rho_{ab}(t)={\rm Tr}_{12}\rho(t)=e^{-iH_{ab}t}\nonumber\\
&\times\sum_{m_a,m_b,n_a,n_b,m_1,m_2=\pm 1} c_{m_a,m_b,m_1,m_2}c^*_{n_a,n_b,m_1,m_2}\vert m_a\ m_b\rangle \langle n_a\ n_b\vert\nonumber\\
&\times\exp{\left[-i\frac{J_0t}{4}(m_a+m_b-n_a-n_b)(m_1+m_2)\right]}\nonumber\\
&\times\exp{\left[-\left(m_a+m_b-n_a-n_b\right)^2\gamma(t)\right]}\nonumber\\
&\times\exp{\left[-i\left((m_a+m_b)^2-(n_a+n_b)^2+2(m_a+m_b-n_a-n_b)(m_1+m_2)\right)\Delta(t)\right]}\nonumber\\
&\times e^{iH_{ab}t}.
\label{spinddensitymatrixfinal3}
\end{align}
Using the explicit form of eigenstates and eigenvalues of the Hamiltonian $H_{ab}$ (see, Appendix~\ref{appegen}), the evolution of the basis states $\vert m_a\ m_b\rangle$ under the operator $e^{-iH_{ab}t}$
in (\ref{spinddensitymatrixfinal3}) takes the form
\begin{eqnarray}
&&e^{-iH_{ab}t}\vert\uparrow\uparrow\rangle=e^{-i\left(J_z/4+h'\right)t}\vert\uparrow\uparrow\rangle,\nonumber\\
&&e^{-iH_{ab}t}\vert\uparrow\downarrow\rangle=e^{iJ_zt/4}(\cos(Jt/2)\vert\uparrow\downarrow\rangle -i \sin(Jt/2)\vert\downarrow\uparrow\rangle),\nonumber\\
&&e^{-iH_{ab}t}\vert\downarrow\uparrow\rangle=e^{iJ_zt/4}(-i\sin(Jt/2)\vert\uparrow\downarrow\rangle + \cos(Jt/2)\vert\downarrow\uparrow\rangle),\nonumber\\
&&e^{-iH_{ab}t}\vert\downarrow\downarrow\rangle=e^{-i\left(J_z/4-h'\right)t}\vert\downarrow\downarrow\rangle.
\label{evolabunderHab}
\end{eqnarray}

\section{Entanglement of the $S_a$ and $S_b$ spins \label{entab}}

We are interested in the entanglement behavior between the $S_a$ and $S_b$ spins. For this purpose, we use negativity as a measure of entanglement \cite{vidal2002, plenio2005}. It is based on the Peres-Horodecki
criterion \cite{peres1996, horodecki1996, zyczkowski1998}. Suppose that the quantum system defined by the density matrix $\rho$ consists of two subsystems $A$ and $B$, respectively.
The Peres-Horodecki criterion confirms that if the partial transpose of the density matrix with respect to subsystem 
$A$($B$) $\rho^{{\rm \Gamma_{A(B)}}}$ has a negative eigenvalue, then the systems $A$ and $B$ are guaranteed to be entangled. For the 2x2 and 2x3 quantum systems, this criterion is necessary and sufficient condition.
Then the negativity of a subsystem $A$($B$) can be expressed as the sum of the negative eigenvalues of $\rho^{{\rm \Gamma_{A(B)}}}$
\begin{eqnarray}
\mathcal{N}(\rho)=\left\vert \sum_{\Lambda_i<0}\Lambda_i\right\vert=\sum_{i}\frac{\vert\Lambda_i\vert-\Lambda_i}{2},
\label{negativity}
\end{eqnarray}
where $\Lambda_i$ are the eigenvalues of the $\rho^{{\rm \Gamma_{A(B)}}}$. Similar to Wootters measure of entanglement \cite{wootters1998,wootters1997}, this quantity allows for an accurate evaluation of the entanglement
in a two-qubit state. Therefore, it can be reliably used to quantify entanglement between two spins. It is also worth noting that, unlike Wootters measure, this measure is more straightforward to apply in practical calculations.

We started with the state where all spins are separated and projected along the positive direction of the $x$-axis. This state can be easily prepared by placing the system in a strong external magnetic field directed along
the $x$-axis. We start with this state because it is completely separable, and we are interested in the behavior of the entanglement in the system that occurs alongside decoherence. Note that with this approach,
other initial states with different levels of entanglement can also be studied, allowing one to observe how maximal entanglement is achieved during the evolution. This state can be expressed as follows
\begin{eqnarray}
\vert\psi_I\rangle_s = \frac{1}{4}\left(\vert\uparrow\uparrow\rangle+\vert\uparrow\downarrow\rangle+\vert\downarrow\uparrow\rangle+\vert\downarrow\downarrow\rangle\right)_{12}\left(\vert\uparrow\uparrow\rangle+\vert\uparrow\downarrow\rangle+\vert\downarrow\uparrow\rangle+\vert\downarrow\downarrow\rangle\right)_{ab}.
\label{instate}
\end{eqnarray}
Substituting the parameters of this states into expression (\ref{spinddensitymatrixfinal3}), we obtain the density matrix that defines the evolution of the $a$ and $b$ spins.
In the base $\vert\uparrow\uparrow\rangle$, $\vert\uparrow\downarrow\rangle$, $\vert\downarrow\uparrow\rangle$, $\vert\downarrow\downarrow\rangle$, it takes the form
\begin{eqnarray}
\rho(t)_{ab}=\left( \begin{array}{ccccc}
\frac{1}{4} & \frac{1}{8}e^{-4z(t)}A(1+B)e^{-ih't} &\\[9pt]
\frac{1}{8}e^{-4z(t)^*}A^*(1+B)e^{ih't} & \frac{1}{4} & \\[9pt]
\frac{1}{8}e^{-4z(t)^*}A^*(1+B)e^{ih't} & \frac{1}{4} &  \\[9pt]
\frac{1}{4}e^{-16\gamma(t)}B^2e^{i2h't} & \frac{1}{8}e^{-4z(t)}A(1+B)e^{ih't}
\end{array}\right.\nonumber\\[9pt]
\left. \begin{array}{ccccc}
\frac{1}{8}e^{-4z(t)}A(1+B)e^{-ih't} & \frac{1}{4}e^{-16\gamma(t)}B^2e^{-i2h't}\\[9pt]
\frac{1}{4} & \frac{1}{8}e^{-4z(t)^*}A^*(1+B)e^{-ih't} \\[9pt]
\frac{1}{4} & \frac{1}{8}e^{-4z(t)^*}A^*(1+B)^{-ih't} \\[9pt]
\frac{1}{8}e^{-4z(t)}A(1+B)e^{ih't} & \frac{1}{4}
\end{array}\right),
\label{densitymatrixabm}
\end{eqnarray}
where we introduce the following notations $z(t)=\gamma(t)+i\Delta(t)$, $A=e^{-i(J_z-J)t/2}$ and $B=\cos\left(J_0t+8\Delta(t)\right)$. In Appendix~\ref{calcnegativity}, we present the derivation of the negativity
for state (\ref{densitymatrixabm}). It takes the form
\begin{align}
&\mathcal{N}=\frac{1}{8}\left\vert 1-e^{-16\gamma(t)}\cos^2\left(J_0t+8\Delta(t)\right)-\left[ \left(1-e^{-16\gamma(t)}\cos^2\left(J_0t+8\Delta(t)\right)\right)^2 \right.\right.\nonumber\\
&\left.\left.+ 16e^{-8\gamma(t)}\sin^2\left((J_z-J)t/2+4\Delta(t)\right)\cos^4\left(J_0t/2+4\Delta(t)\right) \right]^{1/2} \right\vert.
\label{negativityabinstate}
\end{align}
As can be seen, there is a competition between the factor $\gamma(t)$ and $\Delta(t)$.
The factor $\gamma(t)$ causes the decoherence in the system and reduces the entanglement, while the factor $\Delta(t)$ generates an additional interaction between the spins and leads to an increase in entanglement.
The dominance of one factor over the other is defined by the model of the bosonic bath itself. Note that the magnetic field does not affect the entanglement of the system. This is because the magnetic field terms
commute with the other terms in the Hamiltonian and only induce local unitary transformations on each spin. Local unitary transformations do not alter the entanglement of the system.
However, magnetic fields allows one to achieve more states in such a system. Since the magnetic field terms commute with the rest of the Hamiltonian, the system's evolution in the presence of a magnetic field
is no more complex than in its absence. In the absence of a bosonic bath, the entanglement is determined by the interaction coupling between spins \cite{kuzmak20231,kuzmak20232}
\begin{eqnarray}
\mathcal{N}=\frac{1}{8}\left\vert \sin^2\left(J_0t\right)-\left[ \sin^4\left(J_0t\right)+16\sin^2\left((J_z-J)t/2\right)\cos^4\left(J_0t/2\right) \right]^{1/2} \right\vert.
\label{negativityabinstate2}
\end{eqnarray}
Let us consider the behavior of the entanglement between the $S_a$ and $S_b$ spins (\ref{negativityabinstate}) for different cases of the bosonic bath spectral density.

\section{Results for different types of bosonic bath \label{resultssec}}

To evaluate the decoherence parameters \eqref{decfactor}, we use the rule where the summation over the bath modes are replaced by integrals \cite{Morozov2012,Chaudhry2013}
\begin{eqnarray}
\frac{1}{V}\sum_k\vert g_k\vert^2f(\omega_k)=\int_0^{\infty}d\omega J(\omega)f(\omega),
\label{rule}
\end{eqnarray}
where $J(\omega)$ is the spectral density of the bosonic bath. We adopt the most common model of the spectral density used in spin-boson systems (for example, see \cite{Morozov2012,Chaudhry2013,Ignatyuk2022,Barr2024})
\begin{eqnarray}
J(\omega)=\lambda\omega_c^{1-s}\omega^se^{-\omega/\omega_c},
\label{spectraldensity}
\end{eqnarray}
where $\lambda\sim \vert g_k\vert^2$ represents the coupling strength between the spin subsystem and the bosonic bath (it is a dimensionless), $s>0$ is so-called Ohmicity parameter, and $\omega_c$ is the cut-off frequency, indicating
that $J(\omega)\to 0$ as $\omega\to \infty$. Essentially, it defines the frequency range of the boson bath. The value of the parameter $s$ determines different scenarios of spin-environment interaction \cite{Vega2017}.
Depending on the value of $s$, there are three coupling cases: the sub-Ohmic case with $0<s<1$, the Ohmic case with $s=1$, and the super-Ohmic case with $s>1$.
The parameters of the environmental model ($\lambda$, $\omega_c$, and $s$) are selected based on experimental data. For example, the phonon spectrum of azurite has been studied using Raman spectroscopy \cite{azuritespectra}.
This spectrum features several vibrational modes: low-frequency modes in the range of 7.49–16.31 THz, mid-frequency modes from 22.96 to 28.15 THz, and high-frequency modes from 32.89 to 43.74 THz. The Ohmic spectral
density parameters can be chosen to approximate the shape of this spectrum, with the cut-off frequency $\omega_c = 43.74$ THz corresponding to the highest phonon frequencies.
It is also possible to determine the phonon temperature from the Raman spectrum. Raman spectra change with temperature: peak positions shift, linewidths broaden, and intensities vary. In the other hand,
the temperature dependence of the phonon mode intensity is governed by the Bose-Einstein distribution. By analyzing intensity changes at different temperatures, one can extract the phonon inverse-temperature $\beta$.

Thus, decoherence parameters \eqref{decfactor} are expressed by the follows integrals
\begin{eqnarray}
&&\gamma(t)=\frac{\lambda}{4\omega_c^{s-1}}\int_0^\infty  (1-\cos(\omega t))\coth(\beta\omega/2)\omega^{s-2} e^{-\omega/\omega_c}d\omega,\nonumber\\
&&\Delta(t)=\frac{\lambda}{4\omega_c^{s-1}}\int_0^\infty (\sin(\omega t)-\omega t)\omega^{s-2} e^{-\omega/\omega_c}d\omega.
\label{decfactor2}
\end{eqnarray}
It is evident that the magnitude and influence of these parameters on the system's behavior are determined by $\lambda$, which characterizes the strength of the interaction between the spins and the environment,
the model parameter $s$, and the temperature of the environment. It is easy to see that as the temperature increases, $\gamma(t)$ also increases, leading to complete decoherence in the system. The parameter $\lambda$
affects both $\gamma(t)$ and $\Delta(t)$. On one hand, increasing $\lambda$ simultaneously induces decoherence and strengthens the interaction between spins. On the other hand, decreasing $\lambda$
reduces both decoherence and the interaction strength. The influence of parameter $s$ on the system's quantum properties is more complex. Therefore, there exists an optimal set of parameters $\lambda$ and $s$
that maximize the possible entanglement between spins. The asymptotic behavior of the parameter $\gamma(t)$ at large times for different values of $s$ is thoroughly investigated in \cite{Morozov2012}. In Table~I of that paper,
the authors present an analysis of the decoherence parameter $\gamma(t)$ for different environmental regimes. It is worth noting that, within the range $0<s\leq2$, the decoherence parameter increases monotonically,
which enhances decoherence in the system. Conversely, for values of $s>2$, $\gamma(t)$ reaches saturation. The behavior of the $\Delta(t)$ parameter is simpler: its absolute value increases linearly over time,
with the rate of growth depending on $s$. For $s>2$, this rate increases with increasing $s$. In the next subsection, we will illustrate these behaviors with a specific example.

In \cite{Tan2015}, for the case of two spins interacting only with a common environment, the optimal bath parameters to achieve maximal entanglement were estimated numerically. We have obtained an analytical
expression \eqref{negativityabinstate} for the entanglement of spins in the Ising-Heisenberg diamond spin cluster. Next, we analyze the behavior of entanglement for anisotropic and isotropic types of interaction between
$S_a$ and $S_b$ spins with different models of the bosonic environment.

\subsection{Anisotropic interaction between $S_a$ and $S_b$ spins \label{anisotropicint}}

\begin{figure}[!!h]
\centerline{\includegraphics[scale=0.45, angle=0.0, clip]{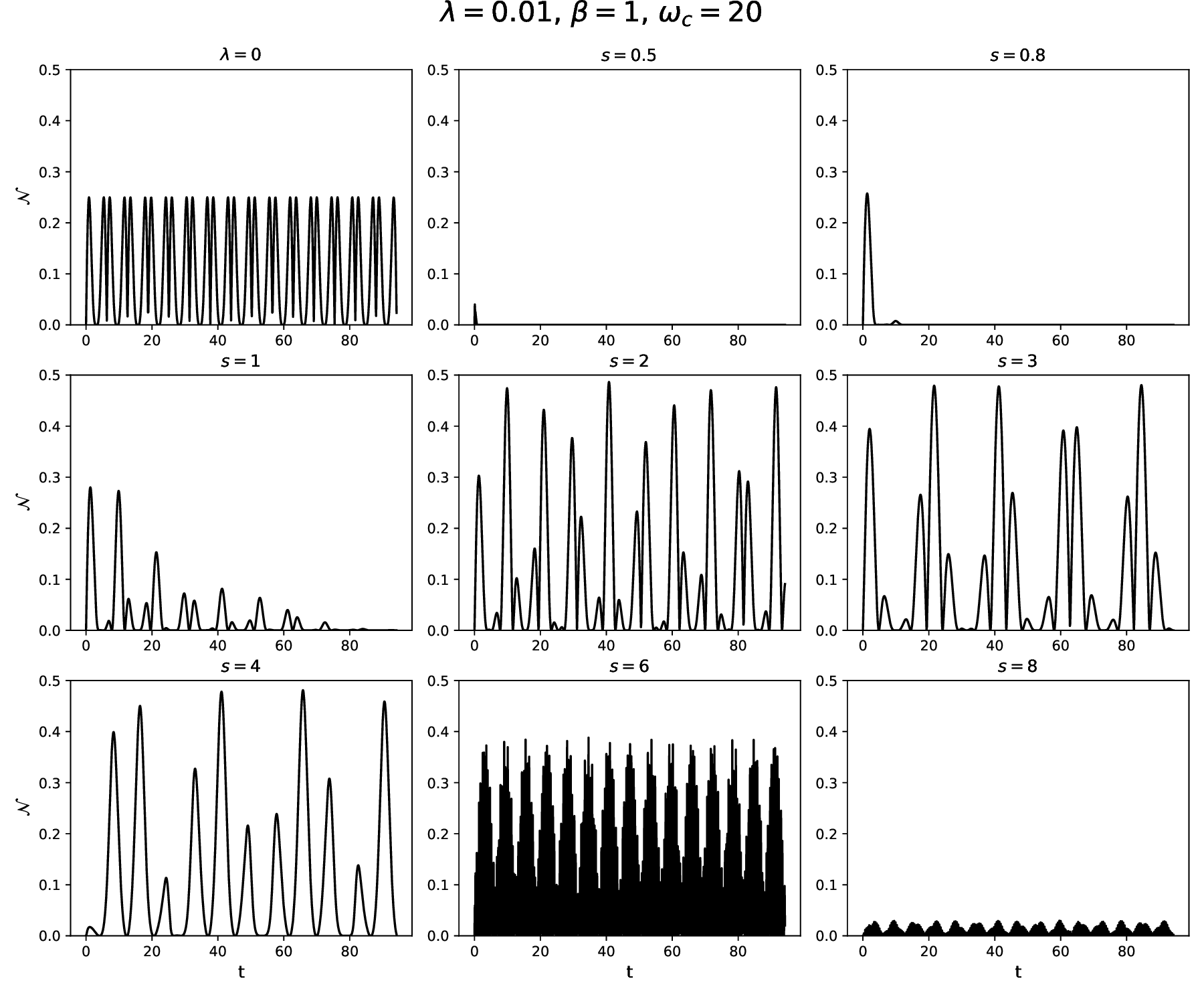}}
\caption{Time dependence of negativity between the $S_a$ and $S_b$ spins in the case of anisotropic Heisenberg interaction with $J=-1$ and $J_z=1$ for different types of environment.
The interaction with the side spins is set to $J_0=1$. The interaction parameter with the bosons subsystem, inverse temperature and the cut-off frequency are $\lambda=0.01$ $\beta=1$ and $\omega_c=20$, respectively.
The first subfigure presents the behavior of negativity without environmental influence ($\lambda=0$). The maximum values of entanglement are taken in the super-Ohmic environment mode for values of $s$ within the values $s\in[2,4]$.}
\label{depNontfordiffs}
\end{figure}

\begin{figure}[!!h]
\centerline{\includegraphics[scale=0.35, angle=0.0, clip]{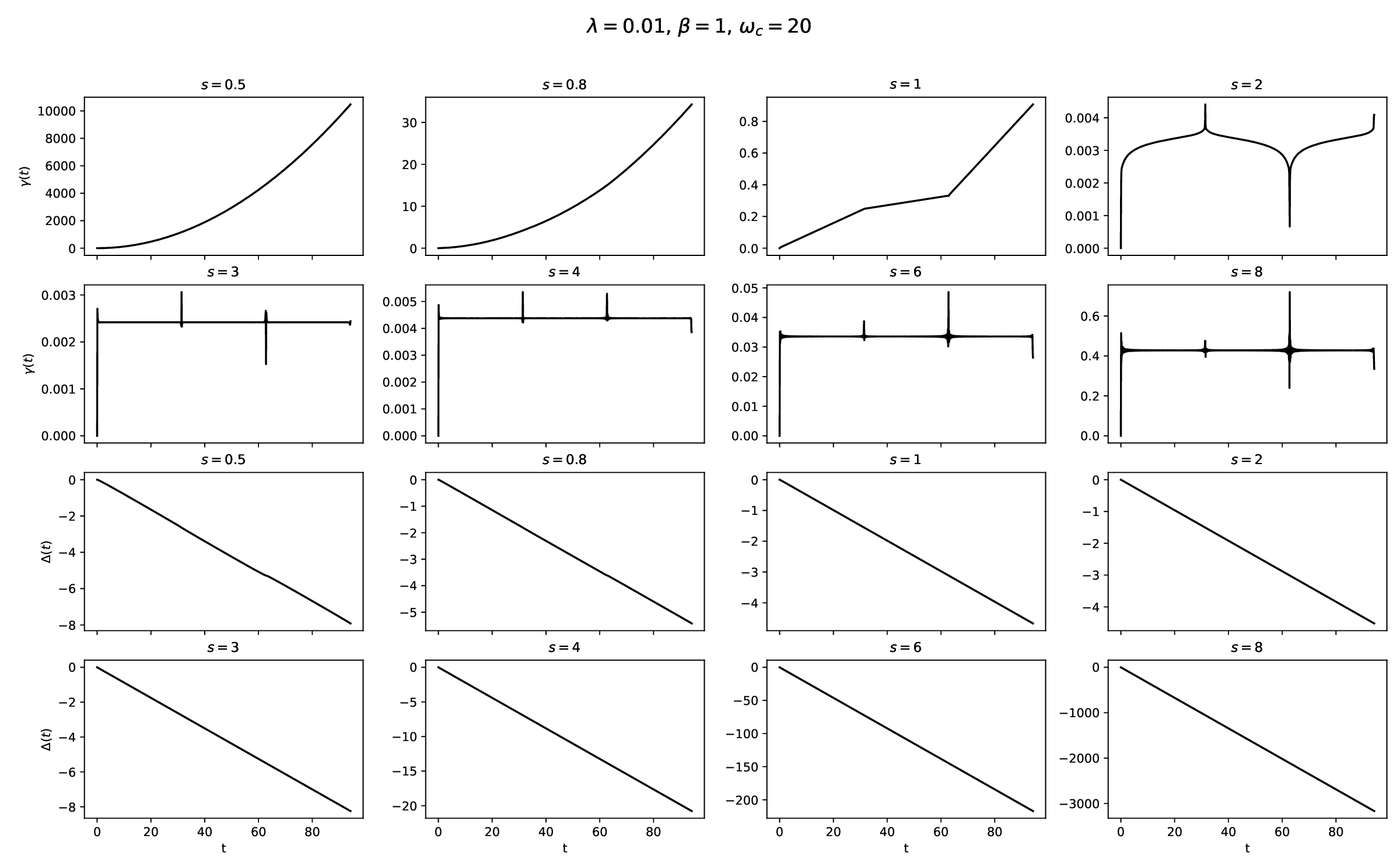}}
\caption{Time dependence of parameters $\gamma(t)$ and $\Delta(t)$. The interaction parameter with the bosons subsystem, inverse temperature and the cut-off frequency are $\lambda=0.01$ $\beta=1$ and $\omega_c=20$, respectively.
The minimal values of $\gamma(t)$ are in the super-Ohmic environment mode for $s$ within the values $s\in[2,4]$. Parameter $\Delta(t)$ has a linear behavior with respect to time, so that the angle
of inclination to the time axis is minimal within the values $s\in[1,3]$.}
\label{gamma_delta}
\end{figure}

\begin{figure}[!!h]
\centerline{\includegraphics[scale=0.45, angle=0.0, clip]{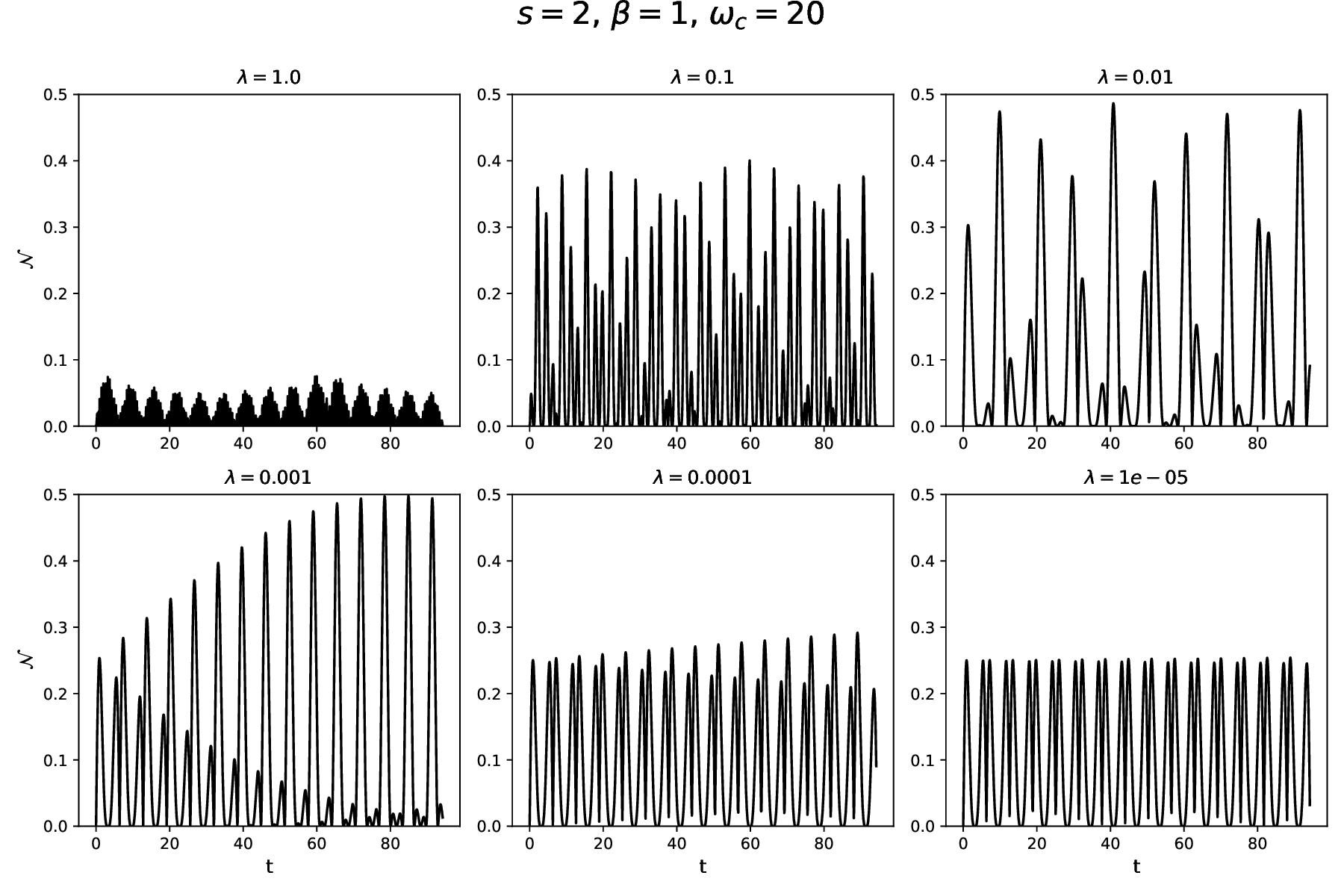}}
\caption{The time dependence of negativity between the $S_a$ and $S_b$ spins in the case of anisotropic Heisenberg interaction is considered for $J=-1$, $J_z=1$ with different values of $\lambda$.
The interaction strength with the side spins is set to $J_0=1$. The inverse temperature and cut-off frequency are $\beta=1$ and $\omega_c=20$, respectively. Here we present results for a super-Ohmic environment with $s=2$.
The maximum values of entanglement are taken when the interaction coupling of spins with the environment is within $\lambda\in[0.001,0.1]$.}
\label{depNontfordifflambda}
\end{figure}

\begin{figure}[!!h]
\centerline{\includegraphics[scale=0.45, angle=0.0, clip]{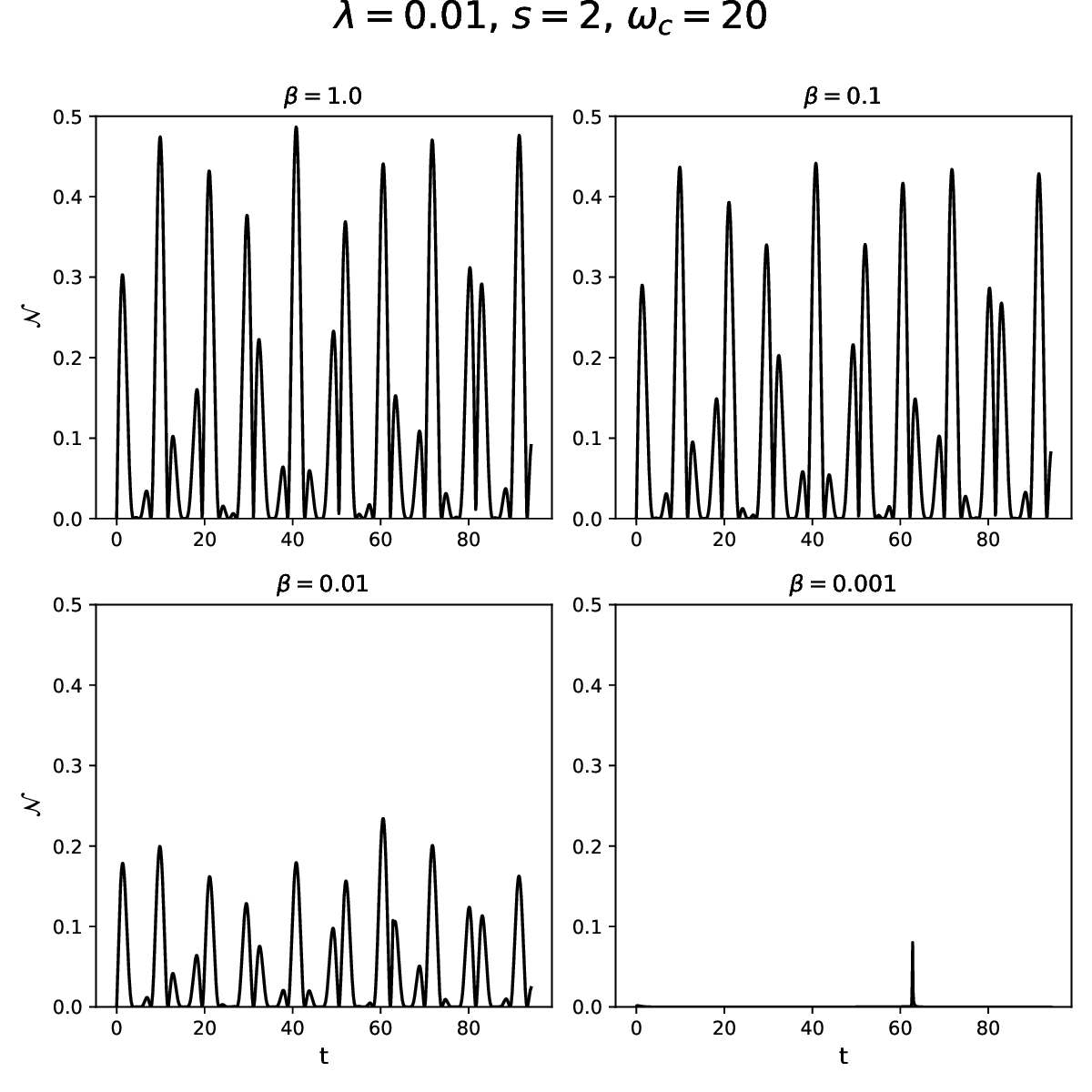}}
\caption{The time dependence of negativity between the $S_a$ and $S_b$ spins in the case of anisotropic Heisenberg interaction with $J=-1$, $J_z=1$ is analyzed with increasing temperature. The interaction strength with the side spins is set to
$J_0=1$. The interaction of each spin with the bosonic bath and the cut-off frequency are $\lambda=0.01$, and $\omega_c=20$, respectively. The results are presented for the super-Ohmic environment with $s=2$.}
\label{depNontfordiffbeta}
\end{figure}

In this subsection, we consider the case where the interaction between the $S_a$ and $S_b$ spins is defined by the anisotropic Heisenberg interaction with parameters $J=-1$ and $J_z=1$. The interaction strength
with the Ising spins is set to $J_0=1$. Consequently, in the absence of an environment, the maximal value of negativity $\mathcal{N}$ between the $S_a$ and $S_b$ spins is achieved. In this case, the time dependence
of negativity is shown in the first subfigure of Fig.~\ref{depNontfordiffs} ($\lambda=0$). First, we investigate the effect of the bosonic bath on the entanglement of spins at temperatures close to zero.
For this purpose, we set $\beta=1$. We assume that the interaction parameter of each spin with the environment is $\lambda=0.01$ and the cut-off frequency is $\omega_c=20$. As shown in Fig.~\ref{depNontfordiffs},
the behavior of negativity varies depending on the parameter $s$. In the sub-Ohmic regime, decoherence dominates, leading to a rapid loss of entanglement. In the Ohmic regime ($s=1$), the system initially reaches
a slightly higher level of entanglement, which subsequently decreases over time. In the super-Ohmic regime, within the range of $s\in[2,4]$, the system periodically reaches nearly the maximum possible entanglement.
However, a further increase in $s$ results in the system losing entanglement. This is because, in this range of $s$, the $\gamma(t)$ function takes on its smallest values. It then increases, leading to a decrease
in entanglement -- though not too rapidly, as the absolute value of the $\Delta(t)$ parameter also increases (Fig.~\ref{gamma_delta}). Additionally note, for larger values of $s$, rapid oscillations in negativity occur.
This is due to the fact that the $\Delta(t)$ parameter appears as an argument of harmonic functions in the expression for negativity and increases rapidly over time when $s$ is large.

Additionally, by adjusting the strength of the bosonic bath interaction with spins for a specific value of $s$, it is possible to increase entanglement. As illustrated in Fig.~\ref{depNontfordifflambda},
for all the selected parameter values and a fixed $s=2$, at $\lambda=0.001$, the system reaches maximally entangled states with $\mathcal{N}=0.5$. As $\lambda$ increases, the influence $\gamma(t)$ becomes dominant
over $\Delta(t)$, and the system experiences stronger decoherence, leading to a loss of entanglement. Conversely, as $\lambda$ decreases, the effect of the environment on the spins weakens, moving the system closer to an isolated state.

Finally, Fig.~\ref{depNontfordiffbeta} demonstrates the system's behavior with increasing temperature. In this scenario, $\gamma(t)$ becomes significantly dominant, and decoherence rapidly leads to entanglement loss.
Therefore, for the implementation of quantum information protocols, it is crucial to maintain such systems at the lowest possible temperatures. Notably, the environment does not always result in decoherence
and the loss of entanglement. With a properly defined set of parameters, it is possible to achieve low decoherence while inducing additional interaction between spins, thereby enhancing entanglement.

\subsection{Isotropic interaction between $S_a$ and $S_b$ spins \label{isotropicint}}

\begin{figure}[!!h]
\centerline{\includegraphics[scale=0.45, angle=0.0, clip]{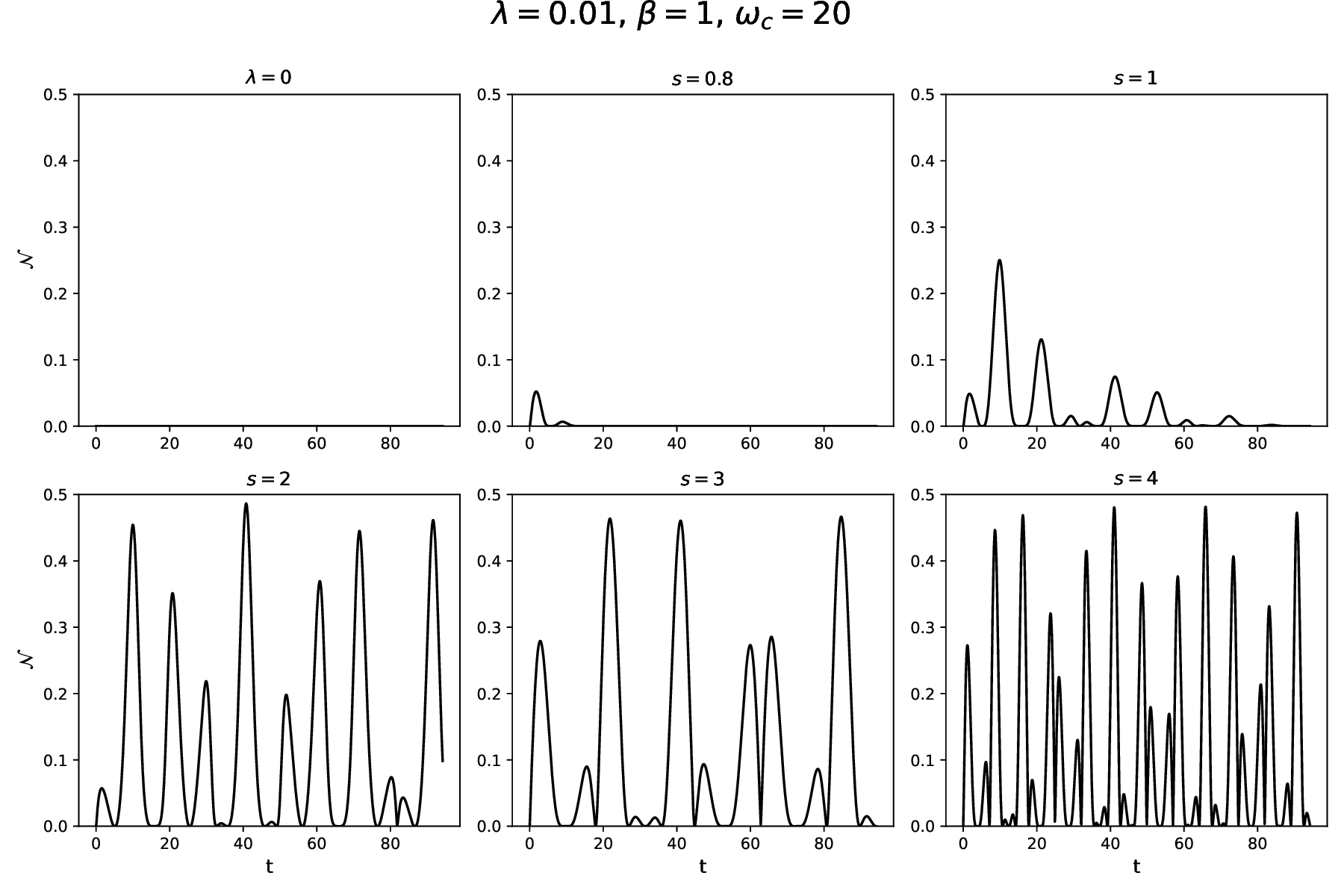}}
\caption{Time dependence of negativity between the $S_a$ and $S_b$ spins in the case of isotropic Heisenberg interaction when $J=1$, $J_z=1$ for different types of environments.
The value of interaction with the side spins is $J_0=1$. The values of the interaction of each spin with the bosonic bath, inverse temperature, and cut-off frequency are $\lambda=0.01$, $\beta=1$ and $\omega_c=20$, respectively.
The first subfigure presents the behavior of negativity without environmental influence ($\lambda=0$). The presence of the environment leads to the emergence of entanglement between the spins.}
\label{isodepNontfordiffs}
\end{figure}

Now, let us suppose that the interaction between the $S_a$ and $S_b$ spins is isotropic, such that $J=J_z$. The main effects arising from the influence of a bosonic bath with different parameters on the entanglement
of the spin system were discussed in the previous subsection. Here, we demonstrate that, in the absence of entanglement within the system, the environment can induce an interaction between spins, which, in turn, leads
to the emergence of entanglement. From equation \eqref{negativityabinstate2}, it follows that, in the absence of a bosonic bath, the entanglement between these spins is zero. However, the interaction of the spins with
the bosonic bath results in the emergence of entanglement within the system. By choosing the same system parameters as in the previous subsection, except for $J=J_z=1$, we illustrate in Fig.~\ref{isodepNontfordiffs}
how entanglement arises between the spins for different values of $s$. In this case, the emergence of entanglement can be attributed to the additional generation of an Ising-type interaction along the $z$-axis.
As a result, the interaction between the spins becomes anisotropic in this direction, which in turn leads to spin entanglement. This effect is also evident in the expression for the negativity (\ref{negativityabinstate}),
where an additional term, $8\Delta(t)$, appears to $J_0t$ and $(J_z-J)t$ terms.

\section{Conclusions \label{conclus}}

One of the main goals of quantum information technology is to identify physical systems capable of efficiently implementing quantum information algorithms. Such systems must be able to generate states that have
quantum entanglement which is a fundamental resource for quantum information processing. One promising candidate is the diamond spin cluster, found in various chemical compounds, including the natural mineral azurite,
where copper ions form a spin diamond chain. In this work, we have considered an Ising-Heisenberg diamond spin cluster, where each spin interacts with a thermal bosonic bath. The cluster consists of two central spins
described by an anisotropic Heisenberg model, interacting with two side Ising spins. The interaction between spins and bosons is described by the dephasing model.

We investigated the time evolution of negativity as a measure of entanglement between the Heisenberg spins, under the influence of a bosonic bath with an Ohmic-type spectral density. Depending on the temperature,
spin-boson coupling strength, and the value of the Ohmicity parameter, the entanglement can either decay or increase. We showed that, for a certain range of environment-determined parameters, entanglement
can be enhanced by the presence of the environment. Furthermore, we demonstrated that in the supe--Ohmic regime at low temperatures, the entanglement between the central spins can reach higher values than
in the absence of the environment. This enhancement is due to the low and saturating decoherence rate in this regime, as well as to the environment-induced effective interaction between spins, which strengthens entanglement.
This effect is particularly evident in the case of isotropic Heisenberg interaction between central spins. In the absence of the environment, the spins do not become entangled. However, due to the dephasing interaction
with the environment, an effective Ising-type coupling is additionally induced. This removes the isotropy of the interaction and leads to the emergence of entanglement between the spins.

\section{Acknowledgements}

The author thanks Prof. Andrij Rovenchak and Dr. Vasyl Ignatyuk for useful comments.

\begin{appendices}

\section{Eigenstates and eigenvalues of the diamond spin cluster \label{appegen}}
\setcounter{equation}{0}
\renewcommand{\theequation}{A\arabic{equation}}

The eigenstates and eigenvalues of the spin subsystem $H_s$ (\ref{spinsystem}) have the following form
\begin{align}
&\vert\psi_1\rangle = \vert\uparrow\uparrow\rangle_{{\tiny 12}}\vert\uparrow\uparrow\rangle_{\small{ab}}, && E_{1}=h+\frac{J_z}{4}+h'+J_0,\nonumber\\
&\vert\psi_2\rangle = \vert\uparrow\uparrow\rangle_{\small{12}}\frac{1}{\sqrt{2}}\left(\vert\uparrow\downarrow\rangle+\vert\downarrow\uparrow\rangle\right)_{ab}, && E_{2}=h+\frac{J}{2}-\frac{J_z}{4},\nonumber\\
&\vert\psi_3\rangle = \vert\uparrow\uparrow\rangle_{\small{12}}\frac{1}{\sqrt{2}}\left(\vert\uparrow\downarrow\rangle-\vert\downarrow\uparrow\rangle\right)_{ab}, && E_{3}=h-\frac{J}{2}-\frac{J_z}{4},\nonumber\\
&\vert\psi_4\rangle = \vert\uparrow\uparrow\rangle_{{\tiny 12}}\vert\downarrow\downarrow\rangle_{\small{ab}}, && E_{4}=h+\frac{J_z}{4}-h'-J_0,\nonumber\\
&\vert\psi_5\rangle = \vert\uparrow\downarrow\rangle_{{\tiny 12}}\vert\uparrow\uparrow\rangle_{\small{ab}}, && E_{5}=\frac{J_z}{4}+h',\nonumber\\
&\vert\psi_6\rangle = \vert\uparrow\downarrow\rangle_{\small{12}}\frac{1}{\sqrt{2}}\left(\vert\uparrow\downarrow\rangle+\vert\downarrow\uparrow\rangle\right)_{ab}, && E_{6}=\frac{J}{2}-\frac{J_z}{4},\nonumber\\
&\vert\psi_7\rangle = \vert\uparrow\downarrow\rangle_{\small{12}}\frac{1}{\sqrt{2}}\left(\vert\uparrow\downarrow\rangle-\vert\downarrow\uparrow\rangle\right)_{ab}, && E_{7}=-\frac{J}{2}-\frac{J_z}{4},\nonumber\\
&\vert\psi_8\rangle = \vert\uparrow\downarrow\rangle_{{\tiny 12}}\vert\downarrow\downarrow\rangle_{\small{ab}}, && E_{8}=\frac{J_z}{4}-h',\nonumber\\
&\vert\psi_9\rangle = \vert\downarrow\uparrow\rangle_{{\tiny 12}}\vert\uparrow\uparrow\rangle_{\small{ab}}, && E_{9}=\frac{J_z}{4}+h',\nonumber\\
&\vert\psi_{10}\rangle = \vert\downarrow\uparrow\rangle_{\small{12}}\frac{1}{\sqrt{2}}\left(\vert\uparrow\downarrow\rangle+\vert\downarrow\uparrow\rangle\right)_{ab}, && E_{10}=\frac{J}{2}-\frac{J_z}{4},\nonumber\\
&\vert\psi_{11}\rangle = \vert\downarrow\uparrow\rangle_{\small{12}}\frac{1}{\sqrt{2}}\left(\vert\uparrow\downarrow\rangle-\vert\downarrow\uparrow\rangle\right)_{ab}, && E_{11}=-\frac{J}{2}-\frac{J_z}{4},\nonumber\\
&\vert\psi_{12}\rangle = \vert\downarrow\uparrow\rangle_{{\tiny 12}}\vert\downarrow\downarrow\rangle_{\small{ab}}, && E_{12}=\frac{J_z}{4}-h',\nonumber\\
&\vert\psi_{13}\rangle = \vert\downarrow\downarrow\rangle_{{\tiny 12}}\vert\uparrow\uparrow\rangle_{\small{ab}}, && E_{13}=-h+\frac{J_z}{4}+h'-J_0,\nonumber\\
&\vert\psi_{14}\rangle = \vert\downarrow\downarrow\rangle_{\small{12}}\frac{1}{\sqrt{2}}\left(\vert\uparrow\downarrow\rangle+\vert\downarrow\uparrow\rangle\right)_{ab}, && E_{14}=-h+\frac{J}{2}-\frac{J_z}{4},\nonumber\\
&\vert\psi_{15}\rangle = \vert\downarrow\downarrow\rangle_{\small{12}}\frac{1}{\sqrt{2}}\left(\vert\uparrow\downarrow\rangle-\vert\downarrow\uparrow\rangle\right)_{ab}, && E_{15}=-h-\frac{J}{2}-\frac{J_z}{4},\nonumber\\
&\vert\psi_{16}\rangle = \vert\downarrow\downarrow\rangle_{{\tiny 12}}\vert\downarrow\downarrow\rangle_{\small{ab}}, && E_{16}=-h+\frac{J_z}{4}-h'+J_0.
\label{eigenvaleigenstate}
\end{align}
The states of the spins are indicated by the subscripts. The states of $S_1$, $S_2$ and $S_a$, $S_b$ spins are denoted by the subscripts $``12"$ and $``ab"$, respectively.

\section{Density matrix \label{densitymatrix}}
\setcounter{equation}{0}
\renewcommand{\theequation}{B\arabic{equation}}
To simplify equation (\ref{evolution1}), we separate the operator $\exp{\left(-i(H_b+H_{sb})t\right)}$. For this purpose, we use Zassenhaus formula \cite{Magnus1954}. The terms of this formula can be obtained
using the code in the Mathematica \cite{Cassas2012}. This formula is well-studied in various paper, so we present it here only up to the third order
\begin{align}
&\exp{\left(-i(H_b+H_{sb})t\right)}=\exp{(-iH_bt)}\exp{(-iH_{sb}t)}\exp{\left(\frac{t^2}{2!}[H_b,H_{sb}]\right)}\nonumber\\
&\times\exp{\left(i\frac{t^3}{3!}([H_b,[H_b,H_{sb}]]+2[H_{sb},[H_b,H_{sb}]])\right)}\ldots 
\end{align}
Now using the explicit form of Hamiltonians $H_b$ (\ref{bath}) and $H_{bs}$ (\ref{spinbath}), and taking into account the commutation relations $[b_k,b_{k'}^+]=\delta_{kk'}$, $[b_k,b_k']=[b_k^+,b_{k'}^+]=0$, we obtain the following result
\begin{align}
&\exp{\left(-i(H_b+H_{sb})t\right)}=\prod_k\exp{\left[-i\omega_kb_k^+b_kt\right]}\nonumber\\
&\times\exp{\left[\frac{1}{\omega_k}(1-\cos(\omega_kt))\left(S_a^z+S_b^z+S_1^z+S_2^z\right)\frac{1}{\sqrt{V}}\left(g_k b_k^+-g_k^*b_k\right)\right]}\nonumber\\
&\times\exp{\left[-i\frac{1}{\omega_k}\sin(\omega_kt)\left(S_a^z+S_b^z+S_1^z+S_2^z\right)\frac{1}{\sqrt{V}}\left(g_k b_k^++g_k^*b_k\right)\right]}\nonumber\\
&\times\exp{\left[i\frac{\vert g_k\vert^2}{V\omega_k^2}\left(\omega_kt-2\sin(\omega_kt)+\sin(\omega_kt)\cos(\omega_kt)\right)\left(S_a^z+S_b^z+S_1^z+S_2^z\right)^2\right]}.
\label{sepoperat}
\end{align}
Next, applying the Baker-Campbell-Hausdorff formula twice
\begin{eqnarray}
&&\exp{(X)}\exp{(Y)}\nonumber\\
&&=\exp{\left(Y+[X,Y]+\frac{1}{2!}[X,[X,Y]]+\frac{1}{3!}[X,[X,[X,Y]]]+\ldots\right)}\exp{(X)},\nonumber
\end{eqnarray}
we replace the operator $\exp{\left[-i\omega_kb_k^+b_kt\right]}$ with other operators and obtain
\begin{align}
&\exp{\left(-i(H_b+H_{sb})t\right)}\nonumber\\
&=\prod_k\exp{\left[\frac{1}{\omega_k}(1-\cos(\omega_kt))\left(S_a^z+S_b^z+S_1^z+S_2^z\right)\frac{1}{\sqrt{V}}\left(g_ke^{-i\omega_kt} b_k^+-g_k^*e^{i\omega_kt}b_k\right)\right]}\nonumber\\
&\times\exp{\left[-i\frac{1}{\omega_k}\sin(\omega_kt)\left(S_a^z+ S_b^z+ S_1^z+ S_2^z\right)\frac{1}{\sqrt{V}}\left(g_ke^{-i\omega_kt} b_k^++g_k^*e^{i\omega_kt}b_k\right)\right]}\nonumber\\
&\times\exp{\left[i\frac{\vert g_k\vert^2}{V\omega_k^2}\left(\omega_kt-2\sin(\omega_kt)+\sin(\omega_kt)\cos(\omega_kt)\right)\left(S_a^z+S_b^z+S_1^z+S_2^z\right)^2\right]}\nonumber\\
&\times\exp{\left[-i\omega_kb_k^+b_kt\right]}.
\end{align}
Substituting this expression into equation (\ref{evolution1}), the time-dependent density matrix takes the form
\begin{align}
&\rho(t)=\frac{1}{Z_b}\prod_k\exp{\left[\frac{1}{\omega_k}(1-\cos(\omega_kt))\left(S_a^z+S_b^z+S_1^z+S_2^z\right)\frac{1}{\sqrt{V}}\left(g_ke^{-i\omega_kt} b_k^+-g_k^*e^{i\omega_kt}b_k\right)\right]}\nonumber\\
&\times\exp{\left[-i\frac{1}{\omega_k}\sin(\omega_kt)\left(S_a^z+S_b^z+S_1^z+S_2^z\right)\frac{1}{\sqrt{V}}\left(g_ke^{-i\omega_kt} b_k^++g_k^*e^{i\omega_kt}b_k\right)\right]}\nonumber\\
&\times\exp{\left[i\frac{\vert g_k\vert^2}{V\omega_k^2}\left(\omega_kt-2\sin(\omega_kt)+\sin(\omega_kt)\cos(\omega_kt)\right)\left( S_a^z+ S_b^z+ S_1^z+S_2^z\right)^2\right]}\nonumber\\
&\times e^{-iH_st}\rho_s(0)e^{iH_st}\nonumber\\
&\times\exp{\left[-i\frac{\vert g_k\vert^2}{V\omega_k^2}\left(\omega_kt-2\sin(\omega_kt)+\sin(\omega_kt)\cos(\omega_kt)\right)\left(S_a^z+S_b^z+S_1^z+S_2^z\right)^2\right]}\nonumber\\
&\times\exp{\left[i\frac{1}{\omega_k}\sin(\omega_kt)\left(S_a^z+S_b^z+S_1^z+S_2^z\right)\frac{1}{\sqrt{V}}\left(g_ke^{-(it+\beta)\omega_k} b_k^++g_k^*e^{(it+\beta)\omega_k}b_k\right)\right]}\nonumber\\
&\times\exp{\left[\frac{1}{\omega_k}(1-\cos(\omega_kt))\left(S_a^z+S_b^z+S_1^z+S_2^z\right)\frac{1}{\sqrt{V}}\left(g_ke^{-(it+\beta)\omega_k} b_k^+-g_k^*e^{(it+\beta)\omega_k}b_k\right)\right]}\nonumber\\
&\times \exp{\left[-\beta\omega_k b_k^+b_k\right]}.
\label{tddensitymatrix}
\end{align}
We observe that due to the spin-boson interaction, an additional Ising interaction between spins emerges, defined by the unitary operator with an effective Hamiltonian proportional to the operator
$\left(S_a^z+S_b^z+S_1^z+S_2^z\right)^2$. We consider the initial state of the spin subsystem in the form
\begin{eqnarray}
\rho_s(0)=\sum_{m_a,m_b,m_1,m_2=\pm 1}\sum_{n_a,n_b,n_1,n_2=\pm 1} c_{m_a,m_b,m_1,m_2}c^*_{n_a,n_b,n_1,n_2}\vert m_a\ m_b\ m_1\ m_2\rangle \langle n_a\ n_b\ n_1\ n_2\vert.
\label{tddensitymatrix2}
\end{eqnarray}
This is the density matrix of an arbitrary pure state of four spins as defined by (\ref{initialstate}). Utilizing the equalities
\begin{eqnarray}
&&\exp{\left[\alpha \left(S_a^z+ S_b^z+ S_1^z+ S_2^z\right)^2\right]}\vert m_a\ m_b\ m_1\ m_2\rangle\nonumber\\
&&=\exp{\left[\frac{\alpha}{4} \left(m_a +m_b+m_1+m_2\right)^2\right]}\vert m_a\ m_b\ m_1\ m_2\rangle,\nonumber\\
&&\exp{\left[A \left(S_a^z+ S_b^z+ S_1^z+ S_2^z\right)\right]}\vert m_a\ m_b\ m_1\ m_2\rangle\nonumber\\
&&=\exp{\left[\frac{A}{4} \left(m_a +m_b+m_1+m_2\right)\right]}\vert m_a\ m_b\ m_1\ m_2\rangle,
\label{equalities}
\end{eqnarray}
where $A$ is the operator which mutually commutes with spin subsystem, and the Weyl's identity
\begin{eqnarray}
e^{A+B}=e^{B}e^{A}e^{[A,B]/2},
\label{weylid}
\end{eqnarray}
where operators $A$ and $B$ mutually commutte with $[A,B]$, we express the density operator in the form
\begin{align}
&\rho(t)=\frac{1}{Z_b}e^{-iH_st}\nonumber\\
&\times\prod_k\sum_{m_a,m_b,m_1,m_2=\pm 1}\sum_{n_a,n_b,n_1,n_2=\pm 1} c_{m_a,m_b,m_1,m_2}c^*_{n_a,n_b,n_1,n_2}\vert m_a\ m_b\ m_1\ m_2\rangle \langle n_a\ n_b\ n_1\ n_2\vert\nonumber\\
&\times\exp{\left[-i\frac{\vert g_k\vert^2}{4V\omega_k^2}(\sin(\omega_k t)-\omega_k t)(\left(\sum_i m_i\right)^2-\left(\sum_in_i\right)^2)\right]}\nonumber\\
&\times\exp{\left[-\frac{\vert g_k\vert^2}{2V\omega_k^2}(1-\cos(\omega_k t))\sinh(\beta\omega_k)\sum_i m_i\sum_i n_i\right]}\nonumber\\
&\times\exp\left[\frac{g_k}{2\sqrt{V}\omega_k}	e^{-i\omega_kt}b_k^+\left(1-\cos(\omega_k t)-i\sin(\omega_k t)\right)\left(\sum_i m_i-e^{-\beta\omega_k}\sum_i n_i\right)\right.\nonumber\\
&\left.-\frac{g^*_k}{2\sqrt{V}\omega_k}e^{i\omega_kt}b_k\left(1-\cos(\omega_k t)+i\sin(\omega_k t)\right)\left(\sum_i m_i-e^{\beta\omega_k}\sum_i n_i\right) \right]\nonumber\\
&\times\exp\left[-\beta\omega_kb_k^+b_k\right]\times e^{iH_st}.
\label{tddensitymatrixfinal}
\end{align}
We then trace out the bosonic subsystem to obtain the state of the spin subsystem, resulting in
\begin{align}
&\rho_s(t)={\rm Tr}_b \rho(t)=e^{-iH_st}\nonumber\\
&\times\prod_k\sum_{m_a,m_b,m_1,m_2=\pm 1}\sum_{n_a,n_b,n_1,n_2=\pm 1} c_{m_a,m_b,m_1,m_2}c^*_{n_a,n_b,n_1,n_2}\vert m_a\ m_b\ m_1\ m_2\rangle \langle n_a\ n_b\ n_1\ n_2\vert\nonumber\\
&\times\exp{\left[-i\frac{\vert g_k\vert^2}{4V\omega_k^2}(\sin(\omega_k t)-\omega_k t)\left(\left(\sum_i m_i\right)^2-\left(\sum_in_i\right)^2\right)\right]}\nonumber\\
&\times\exp{\left[-\frac{\vert g_k\vert^2}{2V\omega_k^2}(1-\cos(\omega_k t))\sinh(\beta\omega_k)\sum_i m_i\sum_i n_i\right]}\nonumber\\
&\times\exp\left[-\frac{\vert g_k\vert^2}{2V\omega_k^2}(1-\cos(\omega_kt))\right.\nonumber\\
&\left.\times\left(\left(\sum_i m_i\right)^2+\left(\sum_i n_i\right)^2 -2\cosh(\beta\omega_k)\sum_i m_i\sum_i n_i\right)\left(\langle b_k^+b_k\rangle+1/2\right) \right]\nonumber\\
&\times e^{iH_st}.
\label{spinddensitymatrix}
\end{align}
Here we use the identity $\left\langle \exp{\left[\gamma b_k^++\alpha b_k\right]} \right\rangle=\exp[\alpha\gamma(\langle b_k^+b_k\rangle+1/2)]$, where $\langle b_k^+b_k\rangle=1/(e^{\beta\omega_k}-1)$.
Finally, after simplification, we arrive at the density matrix of the spin subsystem in the form
\begin{align}
&\rho_s(t)={\rm Tr}_b \rho(t)=e^{-iH_st}\nonumber\\
&\times\sum_{m_a,m_b,m_1,m_2=\pm 1}\sum_{n_a,n_b,n_1,n_2=\pm 1} c_{m_a,m_b,m_1,m_2}c^*_{n_a,n_b,n_1,n_2}\vert m_a\ m_b\ m_1\ m_2\rangle \langle n_a\ n_b\ n_1\ n_2\vert\nonumber\\
&\times\exp{\left[-\left(\sum_i m_i-\sum_i n_i\right)^2\sum_k\frac{\vert g_k\vert^2}{4V\omega_k^2}(1-\cos(\omega_k t))\coth(\beta\omega_k/2)\right]}\nonumber\\
&\times\exp{\left[-i\left(\left(\sum_i m_i\right)^2-\left(\sum_in_i\right)^2\right)\sum_k\frac{\vert g_k\vert^2}{4V\omega_k^2}(\sin(\omega_k t)-\omega_k t)\right]}\times e^{iH_st}.\nonumber\\
\label{spinddensitymatrixfinal}
\end{align}

\section{Calculation of the negativity \label{calcnegativity}}
\setcounter{equation}{0}
\renewcommand{\theequation}{C\arabic{equation}}

Using definition (\ref{negativity}), we calculate the negativity between spins $S_a$ and $S_b$ in the state given by equation (\ref{densitymatrixabm}). The partial transpose of the density matrix $\rho(t)_{ab}$ with respect
to spin $S_b$ has the form
\begin{eqnarray}
\rho(t)_{ab}^{{\rm \Gamma_b}}=\left( \begin{array}{ccccc}
\frac{1}{4} & \frac{1}{8}e^{-4z(t)^*}A^*(1+B)e^{ih't}   &\\[9pt] 
\frac{1}{8}e^{-4z(t)}A(1+B)e^{-ih't}  & \frac{1}{4}    & \\[9pt]
\frac{1}{8}e^{-4z(t)^*}A^*(1+B)e^{ih't} &  \frac{1}{4}e^{-16\gamma(t)}B^2e^{i2h't}  \\[9pt]
\frac{1}{4}   & \frac{1}{8}e^{-4z(t)}A(1+B)e^{ih't}
\end{array}\right.\nonumber\\[9pt]
\left. \begin{array}{ccccc}
\frac{1}{8}e^{-4z(t)}A(1+B)e^{-ih't} & \frac{1}{4} \\[9pt]
\frac{1}{4}e^{-16\gamma(t)}B^2e^{-i2h't}    & \frac{1}{8}e^{-4z(t)^*}A^*(1+B)e^{-ih't} \\[9pt]
\frac{1}{4} & \frac{1}{8}e^{-4z(t)}A(1+B)e^{ih't} \\[9pt]
\frac{1}{8}e^{-4z(t)^*}A^*(1+B)^{-ih't}    & \frac{1}{4}
\end{array}\right).
\label{densitymatrixabmptransposeb}
\end{eqnarray}
From the eigenvalue equation $\det\vert \rho(t)_{ab}^{{\rm \Gamma_b}}-\Lambda I\vert=0$, we obtain the following equations for $\Lambda$
\begin{eqnarray}
&&\Lambda^2-\frac{1}{4}\left(1-e^{-16\gamma(t)}B^2\right)\Lambda+\frac{1}{64}(1+B)^2\left(e^{-4z(t)}A-e^{-4z(t)^*}A^*\right)^2=0,\nonumber\\
&&\Lambda^2-\frac{1}{4}\left(3+e^{-16\gamma(t)}B^2\right)\Lambda+\frac{1}{8}\left(1+e^{-16\gamma(t)}B^2\right)\nonumber\\
&&-\frac{1}{64}(1+B)^2\left(e^{-4z(t)}A+e^{-4z(t)^*}A^*\right)^2=0.
\label{eigenvalueforneg}
\end{eqnarray}
The solutions of these equations are the following
\begin{eqnarray}
&&\Lambda_{1,2}=\frac{1}{8}\left(1-e^{-16\gamma(t)}B^2\right)\nonumber\\
&&\pm\frac{1}{8}\left[\left(1-e^{-16\gamma(t)}B^2\right)^2-(1+B)^2\left(e^{-4z(t)}A-e^{-4z(t)^*}A^*\right)^2\right]^{1/2}\nonumber\\
&&\Lambda_{3,4}=\frac{1}{8}\left(3+e^{-16\gamma(t)}B^2\right)\nonumber\\
&&\pm\frac{1}{8}\left[\left(3+e^{-16\gamma(t)}B^2\right)^2+(1+B)^2\left(e^{-4z(t)}A+e^{-4z(t)^*}A^*\right)^2-8(1+B^2)e^{-16\gamma(t)}\right]^{1/2}.\nonumber\\
\label{eigenvaluesptm}
\end{eqnarray}
It is easy to verify that $\Lambda_1$ is the only negative eigenvalue, and we use it to calculate the negativity given in equation (\ref{negativityabinstate}).

\end{appendices}

{}

\end{document}